\documentclass{epl}

\title{Evidence of the superconducting energy gap in the optical spectra of
$\alpha_t$-(BEDT-TTF)$_{2}$I$_{3}$}
\shorttitle{Superconducting  energy gap of $\alpha_t$-(BEDT-TTF)$_{2}$I$_{3}$}

\author{N. Drichko\thanks{permanent adress: Ioffe Physico-Technical Institute, St. Petersburg, Russia}  \and P. Haas
\and B. Gorshunov\thanks{permanent adress: General Physics Institute, Russian Academy of Sciences, Moscow, Russia} \and D. Schweitzer
\and M. Dressel\thanks{E-mail: \email{dressel@pi1.physik.uni-stuttgart.de}}}
\institute{1. and 3. Physikalisches Institut, Universit\"at Stuttgart - Pfaffenwaldring 57,
D-70550 Stuttgart, Germany}
\shortauthor{N. Drichko \etal}

\begin{document}

\maketitle

\begin{abstract}
The far-infrared in-plane reflectivity of the organic superconductor 
$\alpha_t$-(BEDT-TTF)$_{2}$I$_{3}$ is measured 
down to frequencies of 10~cm$^{-1}$; 
the optical conductivity is obtained by a Kramers-Kronig procedure. 
In the superconducting state an energy gap opens at 25~cm$^{-1}$ which corresponds to $2\Delta/k_BT_c = 4.4$ in agreement with moderate coupling. Even well 
below $T_c=8$~K a considerable absorption remains for $\hbar\omega<2\Delta$. We discuss different posibilities to explain this behaviour.
\end{abstract}

\section{Introduction}
Although organic superconductors have been under intensive investigations for two decades, the mechanism of
superconductivity and the symmetry of the order parameter are still controversial \cite{Ishiguro98}.
Surprisingly, one of the key-experiments for superconductivity --~the optical determination of the
superconducting energy gap $2\Delta$~-- has not been successfully performed yet~\cite{Dressel00}. Soon after the
discovery of superconducting BEDT-TTF salts with a transition temperature 
above 10~K, J. Eldridge and coworkers
found no indications of a gap-related structure in the far-infrared absorption 
spectra  above 10~cm$^{-1}$ by
using a sophisticated bolometric technique \cite{Kornelsen90}. Their results could be understood if the organic
materials were in the clean limit, i.e.,\ $2\Delta>\hbar/\tau$ or $\ell>\xi$, with $\tau$ the scattering time,
$\ell$ the mean free path, $\xi$ the coherence length. This would mean that there is no noticeable spectral weight
change in the range of the gap frequency. Alternatively  the energy gap could 
be very small
compared to the BCS prediction ($2\Delta/h c\approx 25$~cm$^{-1}$ with a typical transition temperature of
BEDT-TTF salts of 10~K) or even missing, either because the order parameter is very anisotropic but still
$s$-wave like or the pairing symmetry is not $s$-wave.

Tunneling spectroscopy  performed on various BEDT-TTF salts 
(mainly the $\kappa$-phase) gives no conclusive
answer. Early reports  claim a large energy gap of $2\Delta/k_BT_c = 9$ and 
a temperature dependence in full
agreement with BCS predictions~\cite{Bando90}. Other results~\cite{Maruyama88,Nomura95a,Nomura95b} vary between
samples but concurrently there seems to be some structure 
around $2\Delta/k_BT_c=3.7$ and 7.4 
which may indicate an
anisotropic energy  gap. Futher investigations still show a inconsistent picture~\cite{Nowack86} as far as the
size and the temperature dependence of the gap are concerned. 
A recent STM study gives evidence of an
anisotropic gap with $d$-wave symmetry and leads to values of $2\Delta/k_BT_c=6.7$ within the highly conducting
planes and to a somewhat larger ratio perpendicular to them~\cite{Arai01}; 
no zero-bias conductance peak is
observed. Various indirect methods
to determine the superconducting energy gap~\cite{proBCS} yield values  in good agreement with the BCS
predictions. However, a considerable number of experimental results point 
to gap nodes and are interpreted as
indications for unconventional superconductivity~\cite{contraBCS}; for recent reviews
see~\cite{Ishiguro98,review}. Since the symmetry of the order parameter and the mechanism of superconductivity in
the two-dimensional organic materials are highly controversial, we reconsidered the most direct method to
investigate the superconducting energy gap by performing optical reflection experiments.

\section{Experimental Results}
\begin{figure}
\twofigures[width=72mm]{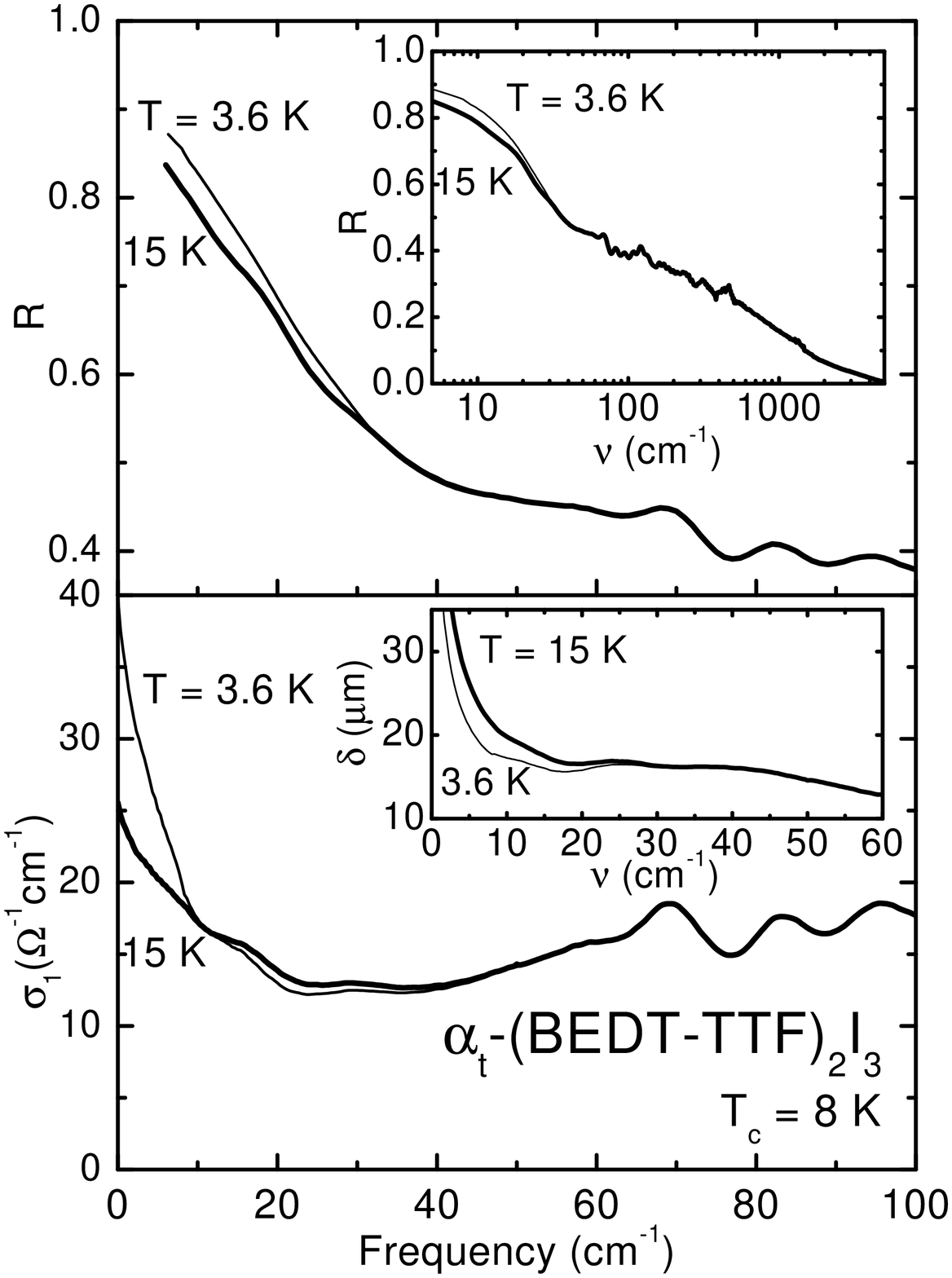}{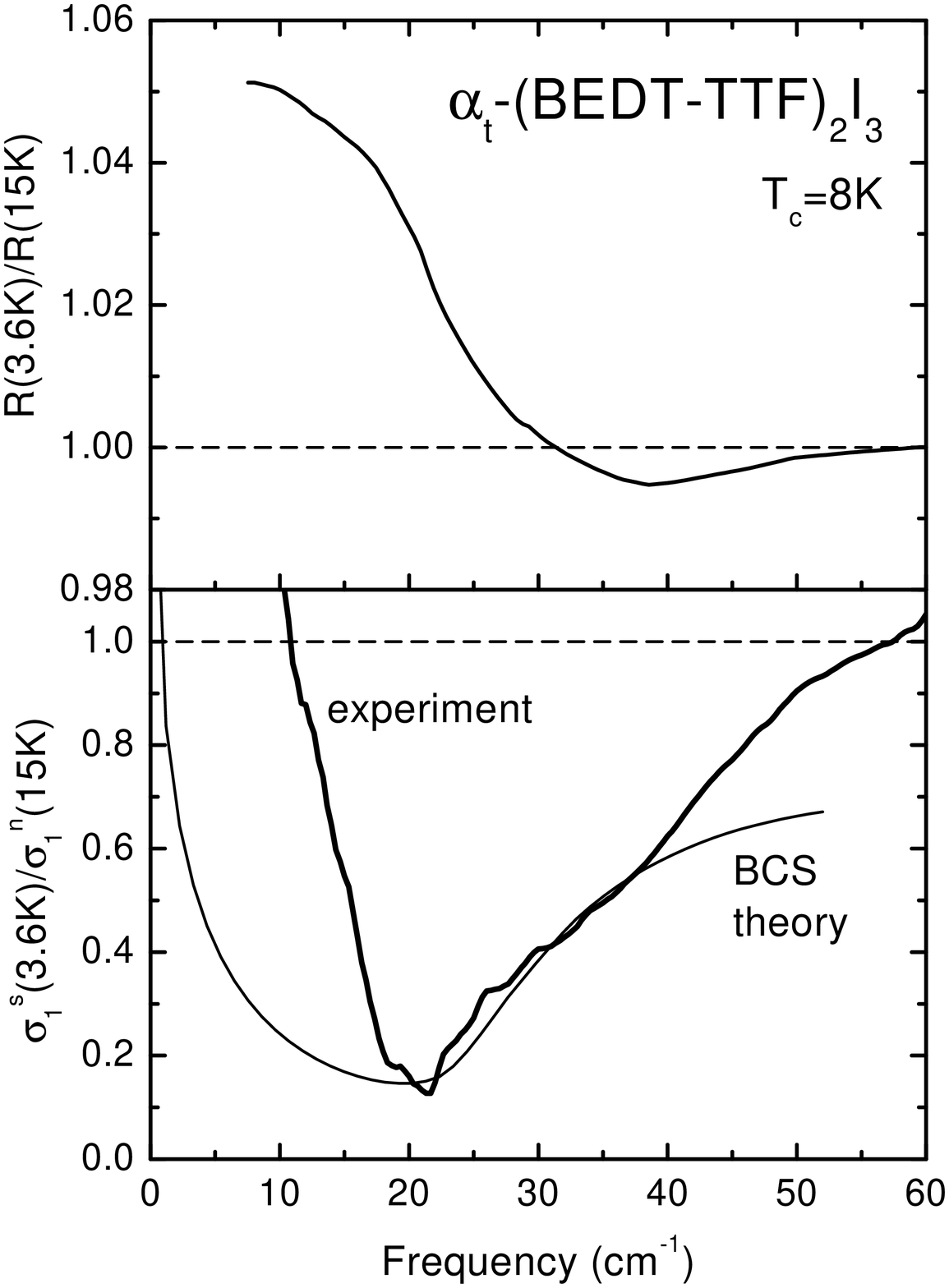} \caption{(a) Frequency dependent reflectivity of
$\alpha_t$-(BEDT-TTF)$_{2}$I$_{3}$ at temperatures above and below the superconducting transition ($T_c=8$~K).
The inset shows the reflectivity at $T=15$~K over a wide spectral range. 
The inset shows the reflectivity  over a
wide frequency range. (b) Corresponding conductivity spectra $\sigma(\omega)$ at $T=15.0$~K and $T=3.6$~K. The generalized 
penetration depth $\delta$ is plotted in the inset as a function of frequency (see text). \label{fig.1}}
\caption{(a) Ratio of the reflectivity spectra of $\alpha_t$-(BEDT-TTF)$_{2}$I$_{3}$ below and above the
superconducting transition: $R(\omega,3.6~{\rm K})/R(\omega,15~{\rm K})$. 
(b)~Frequency dependent conductivity
$\sigma_1^s(\omega,T=3.6~{\rm K})$ in the superconducting state normalized to the normal state conductivity
$\sigma_1^n(\omega,T=15~{\rm K})$ (see text).\label{fig.2}}
\end{figure}

For our investigations we have chosen the $\alpha_t$-(BEDT-TTF)$_2$I$_3$ compound because extremely large
crystals of more than $5\times 3$~mm$^2$ are available which have a high transition temperature of $T_c=8$~K.
Single crystals of $\alpha$-(BEDT-TTF)$_2$I$_3$ were grown by standard electrochemical method and subsequently
transformed to the superconducting $\alpha_t$-phase by termpering at 70$^{\circ}$C over a period of several
hours~\cite{Schweitzer87}. Susceptibility measurements by a SQUID magnetometer indicate that the superconducting
transition has an onset of 8~K but extends below 2~K because grain boundaries remain and internal stress is not
fully released. In agreement with previous experiments the crystals are not completely transformed into the
$\alpha_t$-phase; 
we determine the superconducting volume fraction to be approximately 15\%. 
From the BCS relation
$2\Delta=3.53 k_BT_c$ we expected the gap around 20~cm$^{-1}$, i.e., at the 
lower end of the far-infrared
spectral range. Thus we have utilized a coherent source spectrometer equipped 
with backward wave oscillators
operating in the range from 2 to 45~cm$^{-1}$ to study the reflectivity 
within the highly conducting
plane. A home-made cryostat allowed to go to tempertures as low as 2~K. 
In order to increase the sensitivity for
the comparably small changes of the reflection $R$ upon
the superconducting transition, 
we measured the ratio of
$R(T<T_c)$ to $R(T>T_c)$ without moving the sample. The absolute value of $R(\omega)$ at a certain temperature $T$
was then obtained by replacing the sample to a aluminum mirror of known reflectivity. 
The data were completed by
standard reflection measurements using a Fourier-transform infrared spectrometer 
(inset of Fig.~\ref{fig.1}a).
The low-frequency results at temperatures well above ($T=15.0$~K) and below ($T=3.6$~K) the superconducting
transition are plotted in Fig.~\ref{fig.1}. After extrapolating to $\omega=0$ with a 
Hagen-Rubens behaviour and
to higher frequencies utilizing published data \cite{Zelezny90}, we performed a 
Kramers-Kronig analysis 
to obtain the optical conductivity at different temperatures (Fig.~\ref{fig.1}b). Although
small, the changes in the reflectivity and in the real part 
of the conductivity $\sigma_1(\omega)$ are reliably detected in our
measurements: 
in the super\-conducting state the reflectivity increases and the conductivity 
decreases below approximately 40~cm$^{-1}$. 
Similar observations 
have been made on several crystals.
The effects are better seen in Fig.~\ref{fig.2} where the ratios are plotted: 
upon entering the
super\-conducting state a significant increase in reflectivity is observed 
below 30~cm$^{-1}$ with some undershot
up to approximately 60~cm$^{-1}$. 
The transition to the superconducting state is also unambiguously seen in frequency
dependent penetration depth of the electromagnetic radiation (inset in
Fig.~\ref{fig.1}b) calculated by the general formula 
$\delta =c/(\omega k)$ where $k$ is the extinction coefficient \cite{DresselGruner02}. 
In a BCS superconductor
$\delta(\omega)$ behaves Drude-like ($\delta \propto \omega ^{-0.5}$) 
above the gap frequency, but decreases
considerably  and gets dispersionless for smaller frequencies 
\cite{Tinkham96,Timusk99}. In our case only a slight decreasing and 
flattening of $\delta(\omega)$ is seen in the
superconducting state, connected with the additional absorption processes 
which will be discussed below. 

\section{Analysis and Discussion}
For our further anaylsis we substracted the phonon contributions and 
mid-infrared bands from the conductivity
spectra by fitting them with Lorentzians since only the free electrons 
are affected by the superconducting
transition. In Fig.~\ref{fig.2}b the ratio of the superconducting state and 
the normal state conductivity is
plotted: $\sigma_1^s(\omega,T=3.6~{\rm K})/\sigma_1^n(\omega,T=15~{\rm K})$ 
taken the effective superconduting
fraction into account \cite{Stroud75}. We clearly see a dip in the conductivity 
spectrum as predicted by the BCS
theory~\cite{Tinkham96,DresselGruner02}. 
In a first approximation the minimum  corresponds to the
superconducting energy gap. Thus for $\alpha_t$-(BEDT-TTF)$_{2}$I$_{3}$ 
we evaluate $2\Delta/hc = (25\pm
3)$~cm$^{-1}$ or $2\Delta/k_BT_c=4.4\pm 0.5$ in good agreement with 
what is expected for a medium to strong
coupling BCS superconductor. We cannot make any profound statement 
on the temperature dependence of the gap
frequency.

From the BCS weak-coupling model of an $s$-wave order parameter, 
at $T=0$ there should be no absorption for
energies below $2\Delta$ because the photon energy is not sufficient to break 
the Cooper pairs, implying that
$\sigma_1^s=0$. With increasing temperature the $\omega=0$ peak broadens and 
the gap gradually fills in due to
thermal excitations of single-particles; only above $T/T_c\approx 0.5$ the gap 
starts to decrease noticeably
according to the mean-field behaviour~\cite{Tinkham96,DresselGruner02}. 
Most remarkably our results show a very
broad $\omega=0$ absorption peak which essentially extends all the way 
to the gap frequency. We thus conclude that
there remains some additional absorption channel even at temperatures 
considerably below $T_c$ connected with a
low superconducting plasma frequency
and large London penetration depth. We discuss three possibilities:\\
(i) The absorption  in the superconducting phase might be due to intrinsic states 
in the energy gap. Similar
observations are reported for the high-temperature superconductors 
where nodes in the gap are expected due to
$d$-wave symmetry of the order parameter~\cite{Timusk99}. 
Calculations of the conductivity spectra using models
with an anisotropic energy gap very much depend on the 
actual ${\bf k}$-dependence of the gap and whether there
are nodes present or not \cite{Graf95}. In any case, at low temperature 
the absorption gradually increases
starting from $\omega=0$ to a maximum gap frequency $\Delta_{\rm max}/\hbar$.  Qualitatively simlar behaviour is
observed for an anisotropic gap or some $s+d$ symmetry which results in 
a vanishing absorption below some minimum
energy gap $\Delta_{\rm min}$. Hence,
these explainations are clearly in contrast to our observations.\\
(ii) If large parts of the sample are still in the normal state, we expect a background conductivity up to the
frequency of the scattering rate. From Fig.~\ref{fig.1}b 
we see that the normal state conductivity
contains a rather broad Drude component of $1/(2\pi c\tau_1)\approx 
200$~cm$^{-1}$ width
and in addition a narrower Drude
mode with $1/(2\pi c\tau_2)\approx 10$~cm$^{-1}$. 
Similar observations have been made in most organic conductors
\cite{Dressel00}. The observed behaviour could be explained using a two-band 
model where only one band contributes
to superconductivity. The actual bandstructure of 
$\alpha_t$-(BEDT-TTF)$_{2}$I$_{3}$ is not know. 
While in the
original $\alpha$-phase in general two bands cross the Fermi energy, 
the structure of the $\beta$-phase which is
assumed to be close to that of the
$\alpha_t$-compound has only one cylindrical Fermi surface \cite{Wosnitza96}. 
For some superconducting
heavy fermions it was suggested  that only the electrons at one part 
of the Fermi surface become superconducting  \cite{Knoepfle96}, 
nevertheless such a scenario seems very unlikely in the present case.\\
(iii) The additional below-gap absorption in the superconducting state 
might be due to strong  influence  of the
low frequency phonons and could be treated in the framework of the 
Eliashberg theory. Such treatment allows, in
addition to a better fit of the detected low-energy absorption, also to 
describe the high-frequency wing of the gap feature
where the changes upon entering the superconducting state are confined to 
about twice the gap energy, in contrast
to the usual BCS behaviour. Since the actual low-frequency phonon spectrum 
is not known for this compound we restrained
ourselves from performing a detailed fit.
We want to note, however, that resonant Raman experiments on 
$\alpha_t$-(BEDT-TTF)$_{2}$I$_{3}$ show a vanishing of low-energy phonon bands
at 32 and 42~cm$^{-1}$ below $T_c$ \cite{Ludwig95}.\\
Thus some alternative mechanism for the absorption seems to be relevant.

The changes in the conductivity spectra in the superconducting state are 
rather small 
(as
compared to conventional or high-$T_c$ superconductors), 
leading to a correspondingly small spectral weight of
the superconducting condensate ($\delta$-peak). 
This spectral weight can be estimated from the missing area in
the conductivity spectrum which we calculate as the difference between 
the area under the $\sigma_1^n(T=15$~K)
and the $\sigma_1^s(T=3.6$~K) spectra (starting with the frequency 10~cm$^{-1}$). 
Thus we get the plasma frequency
of the condensate $\omega^{s}_p\geq 25~{\rm cm}^{-1}$, and also estimate correspondingly the London penetration
depth as $\lambda_L=c/\omega^{s}_p\leq 6~\mu$m. 
Magnetization measurements gave 
$\lambda_L\approx 430$~nm~\cite{Gogu88}  for the in-plane penetration depth. 
Our values of $\omega^{s}_p$ and $\lambda_L$ contain some uncertainty 
since we had to make assumptions on  the reflectivity and, correspondingly, the conductivity below 10~cm$^{-1}$. 
Nevertheless, our results on $\lambda_L$ are in good agreement with reports of a 
few $\mu$m for other BEDT-TTF salts~\cite{Ishiguro98,Dressel00}.

It remains to be seen whether these findings can be generalized to other organic superconductors. In particular
the $\kappa$-phase of the BEDT-TTF salts would be of high interest 
because for these compounds the wealth of
informations has been accumulated over the years. 
The corresponding optical experiments are in progress.

\section{Conclusion}
We performed in-plane optical reflection experiments on the organic superconductor $\alpha_t$-(BEDT-TTF)$_{2}$I$_{3}$
and determined an energy gap of 25~cm$^{-1}$ 
which opens in the superconducting phase below $T_c=8$~K  and
corresponds to $2\Delta/k_BT_c=4.4$ (moderate coupling). 
The strong low-frequency absorption which
remains in the superconducting state cannot be fully explained yet. 
The London penetration depth $\lambda_L$ is
estimated as 6~$\mu$m.

\acknowledgments We thank W. Schmidt for assistance during sample preparation and O.Dolgov for useful
discussions. N.D. acknowledges support by the DAAD.

\end{document}